\documentclass[%
 aps, 
 amsmath,amssymb,
preprint,%
]{revtex4-2}

\usepackage{graphicx}
\usepackage{dcolumn}
\usepackage{bm}

\usepackage[utf8]{inputenc}
\usepackage[T1]{fontenc}
\usepackage{mathptmx}
\usepackage{etoolbox}
\usepackage{color}

\makeatletter
\def\@email#1#2{%
 \endgroup
 \patchcmd{\titleblock@produce}
  {\frontmatter@RRAPformat}
  {\frontmatter@RRAPformat{\produce@RRAP{*#1\href{mailto:#2}{#2}}}\frontmatter@RRAPformat}
  {}{}
}%
\makeatother

\makeatletter
\DeclareRobustCommand{\iscircle}{\mathord{\mathpalette\is@circle\relax}}
\newcommand\is@circle[2]{%
  \begingroup
  \sbox\z@{\raisebox{\depth}{$\m@th#1\bigcirc$}}%
  \sbox\tw@{$#1\square$}%
  \resizebox{!}{\ht\tw@}{\usebox{\z@}}%
  \endgroup
}
\makeatother
\begin{document}


\title[Optical creation and annihilation of skyrmion patches in a bulk chiral magnet]{Optical creation and annihilation of skyrmion patches in a bulk chiral magnet}
\author{J. Kalin$^*$}\email{jantje.kalin@ptb.de}
\affiliation{Physikalisch-Technische Bundesanstalt, 38116 Braunschweig, Germany}

\author{D. Kalin}
\affiliation{ESE Engineering und Software-Entwicklung GmbH, 38122 Braunschweig, Germany}

\author{S. Sievers}%
\affiliation{Physikalisch-Technische Bundesanstalt, 38116 Braunschweig, Germany}

\author{A. Bauer}
\affiliation{Physik-Department, Technische Universität München, 85748 Garching, Germany}
\affiliation{Zentrum für QuantumEngineering (ZQE), Technische Universität München, 85748 Garching, Germany}

\author{H. W. Schumacher}
\affiliation{Physikalisch-Technische Bundesanstalt, 38116 Braunschweig, Germany}

\author{R. Abram}
\affiliation{Physikalisch-Technische Bundesanstalt, 38116 Braunschweig, Germany}

\author{H. Füser}
\affiliation{Physikalisch-Technische Bundesanstalt, 38116 Braunschweig, Germany}

\author{C. Pfleiderer}
\affiliation{Physik-Department, Technische Universität München, 85748 Garching, Germany}
\affiliation{Zentrum für QuantumEngineering (ZQE), Technische Universität München, 85748 Garching, Germany}
\affiliation{Munich Center for Quantum Science and Technology (MCQST), Technische Universität München, 85748 Garching, Germany}

\author{M. Bieler}
\affiliation{Physikalisch-Technische Bundesanstalt, 38116 Braunschweig, Germany}

\date{\today}

\begin{abstract}
A key challenge for the realization of future skyrmion devices comprises the controlled creation, annihilation and detection of these topologically non-trivial magnetic spin textures. In this study, we report an all-optical approach for writing, deleting, and reading skyrmions in the cubic chiral magnet Fe$_{0.25}$Co$_{0.75}$Si based on thermal quenching. Using focused femtosecond laser pulses, patches of a skyrmion state are created and annihilated locally, demonstrating unprecedented control of thermally metastable skyrmions in a bulk compound. The skyrmion state is read-out by analyzing the microwave spin excitations in time-resolved magneto-optical Kerr effect measurements. Extracting the magnetic field and laser fluence dependence, we find well-separated magnetic field regimes and different laser fluence thresholds for the laser-induced creation and annihilation of metastable skyrmions. The all-optical skyrmion control, as established in this study for a model system, represents a promising and energy-efficient approach for the realization of skyrmions as magnetic bits in future storage devices, reminiscent of magneto-optical storage devices in the past. 

\end{abstract}
\maketitle
The formation of skyrmions, nanometer-sized magnetic whirls, from a topologically trivial state requires topological winding to build up the skyrmion's inherent spin structure. Consequently, once formed skyrmions tend to be remarkably robust, characterized by high topological energy barrier and long characteristic time scale for skyrmion unwinding. In cubic chiral magnets, skyrmions are typically observed as an equilibrium state within a narrow parameter range of the magnetic phase diagram at temperatures just below the onset of long-ranged magnetic order.
However, the large characteristic time scales for skyrmion unwinding allows to create a metastable skyrmion state across larger parts of the phase diagram by thermal quenching \cite{okamura2016transition,munzer2010skyrmion,bauer2018skyrmion,oike2016interplay,bannenberg2016extended,takagi2021hybridized,seki2020propagation, kalin2022optically, milde2013unwinding}.
Also in ferromagnetic multilayer systems, where skyrmions may often not emerge as equilibrium states, a metastable skyrmion state can be induced \cite{hrabec2017current, legrand2017room, buttner2017field, lemesh2018current,zhang2018direct, temiryazev2018formation, qin2018size}. Most notably, skyrmions may be locally created in such multilayer systems by the irradiation with multiple \cite{finazzi2013laser} or even single femtosecond (fs) laser pulses \cite{gerlinger2021application, je2018creation,kern2022tailoring}. 
Thus, the focused laser light triggers topological winding in this material class, which is commonly explained by lowering the topological energy barrier by laser excitation \cite{gerlinger2021application, buttner2017field}. 
So far, the laser-induced creation of skyrmions has only been demonstrated in multilayer and thin-film systems \cite{finazzi2013laser,je2018creation,gerlinger2021application,kern2022tailoring,kern2022deterministic,buttner2021observation}. In bulk compounds of cubic chiral magnets no laser writing of skyrmions has been achieved to date. Furthermore, the inverse process, laser-induced skyrmion annihilation, is largely unexplored
and laser writing experiments have not yet been combined with optical readout techniques. However, this presents the possibility of fast and energy-efficient all-optical skyrmion devices, where the operations of writing, deleting, and reading can be accomplished using optical means.

In this Letter, we explore the all-optical control of the metastable skyrmion lattice state (MSkL) in a bulk sample of the cubic chiral magnet Fe$_{0.25}$Co$_{0.75}$Si.
By reading out the magnetic state in time-resolved magneto-optical Kerr effect (TR-MOKE) measurements, we systematically track the creation and annihilation of MSkL patches after the irradiation with multiple fs laser pulses and subsequent cryogenic cooling. We find laser-induced MSkL creation and annihilation in distinct and well-separated magnetic field regimes for different laser fluence thresholds, reflecting the distinct
temperature regimes for the stabilization and destabilization of the MSkL. Overall, this study demonstrates the feasibility of all-optical skyrmion read-write-delete operations within a single experimental framework.

\begin{figure}
\includegraphics{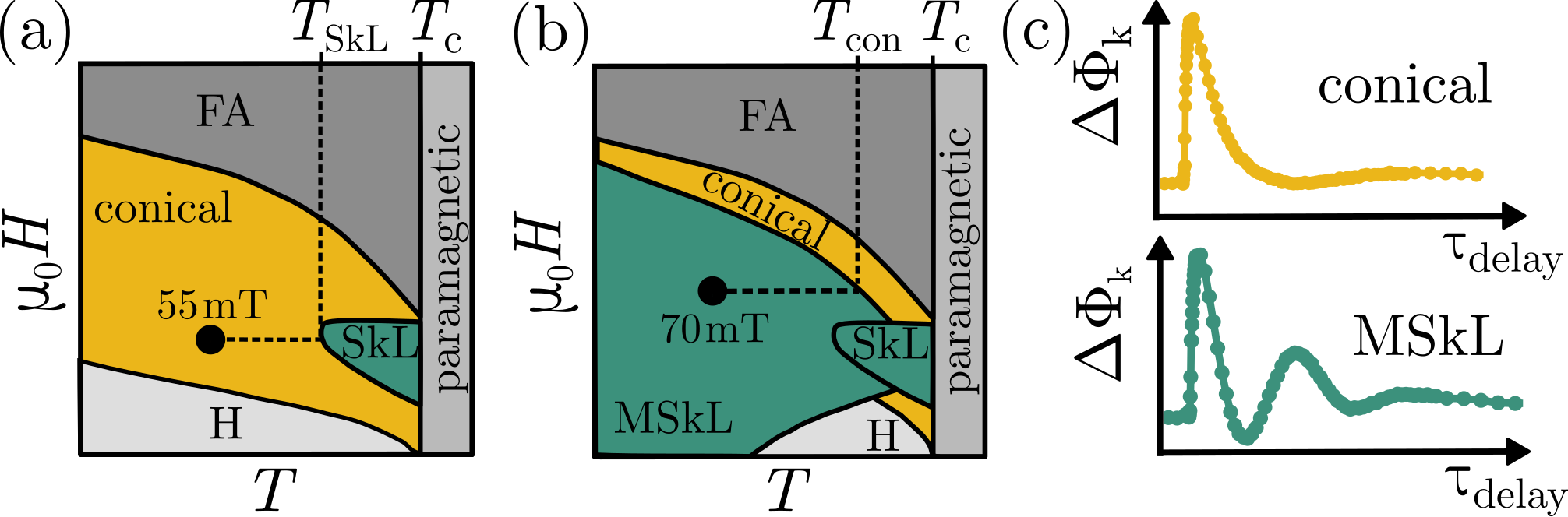}
\caption{\label{fig:Figure1} Schematic magnetic phase diagram of Fe$_{0.25}$Co$_{0.75}$Si (a) in thermal equilibrium and (a) following thermal quenching through the SkL. (c) Characteristic out-of-plane magnetization dynamics of the equilibrium conical state (top panel) and the MSkL (bottom panel) measured in TR-MOKE experiments as Kerr angle change ($\Delta \phi_k$) as a function of the time delay between pump and probe beam $\tau_{\text{delay}}$.
}
\end{figure}

Fe$_{0.25}$Co$_{0.75}$Si exhibits the generic magnetic phase diagram of cubic chiral magnets \cite{munzer2010skyrmion}, displayed in Fig. \ref{fig:Figure1}(a). While being a paramagnet for temperatures above the ordering temperature ${T_{\text{c}} =39\,\text{K}}$, magnetic order sets in for $T<T_{\text{c}}$ comprising long-range helimagnetism (H), conical order, and a field-aligned (FA) state under increasing magnetic field \cite{bauer2016history}. Most importantly, in a limited temperature and magnetic field range just below $T_{\text{c}}$, a hexagonal lattice of skyrmions \cite{yu2010real} is observed, which we refer to as an equilibrium skyrmion lattice state (SkL). 
Previous studies established that thermal quenching under magnetic fields crossing the SkL may avoid the equilibrium phase transition between the skyrmion lattice and conical state kinetically, resulting in a MSkL extending over large parts of the magnetic phase diagram, see Fig. \ref{fig:Figure1}(b). The lifetime of the MSkL, representing the characteristic time scale for MSkL unwinding, is exponentially dependent on temperature and significantly influenced by the applied magnetic field \cite{wild2017entropy}. 
When the magnetic field is outside the SkL range, an increase in temperature gradually induces the decay of the MSkL through skyrmion unwinding on experimentally relevant time scales. This thermal destabilization of the MSkL results in the full transition to the conical state at $T_\text{con}$, see Fig. \ref{fig:Figure1}(b). So far, for cubic chiral magnets MSkL was realized by thermal quenching of the entire sample
\cite{bannenberg2016extended, munzer2010skyrmion, seki2020propagation,oike2016interplay,berruto2018laser}, leading to
a homogenous MSkL that extends over the whole crystal.
This was achieved through techniques such as rapid field cooling  \cite{bannenberg2016extended, munzer2010skyrmion, seki2020propagation}, where the sample is cooled at high cooling rates from temperatures above $T_{\text{c}}$ to lower temperatures using cryogenic temperature control. When magnetic fields outside the SkL field regime are applied during this cooldown, the equilibrium conical state forms. However, for magnetic fields within the SkL, the rapid cooling results in the formation of the MSkL.

In a recent study \cite{kalin2022optically}, it was established that the TR-MOKE measurements of out-of-plane magnetization dynamics serve a sensitive probe for MSkL detection. While the equilibrium conical state, see upper panel in Fig. \ref{fig:Figure1}(c), exhibit a non-oscillatory de- and remagnetization characteristic, in the MSkL, under the same magnetic field and temperature, the skyrmion breathing mode \cite{mochizuki2012spin, onose2012observation, schwarze2015universal} is observed as a precessional signal, see lower panel in Fig. \ref{fig:Figure1}(c).
In this work, we adopt this approach to study laser-induced MSkL creation and annihilation by performing TR-MOKE measurements in a pump-probe experiment. To that end, we apply laser pulses of about ${\tau_{\text{pulse}}=150\,\text{fs}}$ duration with photon wavelength of ${\lambda=800\,\text{nm}}$ and a repetition rate of ${G=76\,\text{MHz}}$. The concentrical pump pulses with focal $1/e^2$ radius of ${w_L=60\,\mu\text{m}}$ and probe pulses with focal $1/e^2$ radius of ${30\,\mu\text{m}}$ are focused under near normal incidence onto the surface of the sample. The latter is located in an optical cryostat with the magnetic field pointing out-of the sample plane. The linearly polarized pump beam thermally triggers the magnetization dynamics of the sample. The dynamics is detected by the polarization change of the reflected probe beam as induced by the magneto-optical Kerr effect (MOKE), i.e., by measuring the Kerr rotation $\Delta \phi_k$ in a stroboscopic manner as a function of the time delay between probe and pump beam $\tau_{\text{delay}}$. The measurement technique is discussed in more detail in Ref. \cite{kalin2022optically}. In this study, the laser fluence is quantified by the peak fluence of the Gaussian laser beam profile absorbed by the sample. For a non-invasive detection of the magnetic states we choose the lowest possible pump laser fluence for the TR-MOKE measurements ${F=2.2\,\mu\text{Jcm}^{-2}}$ to limit the temperature increase by steady state heating to about ${5\,\text{K}}$.

\begin{figure}
\includegraphics{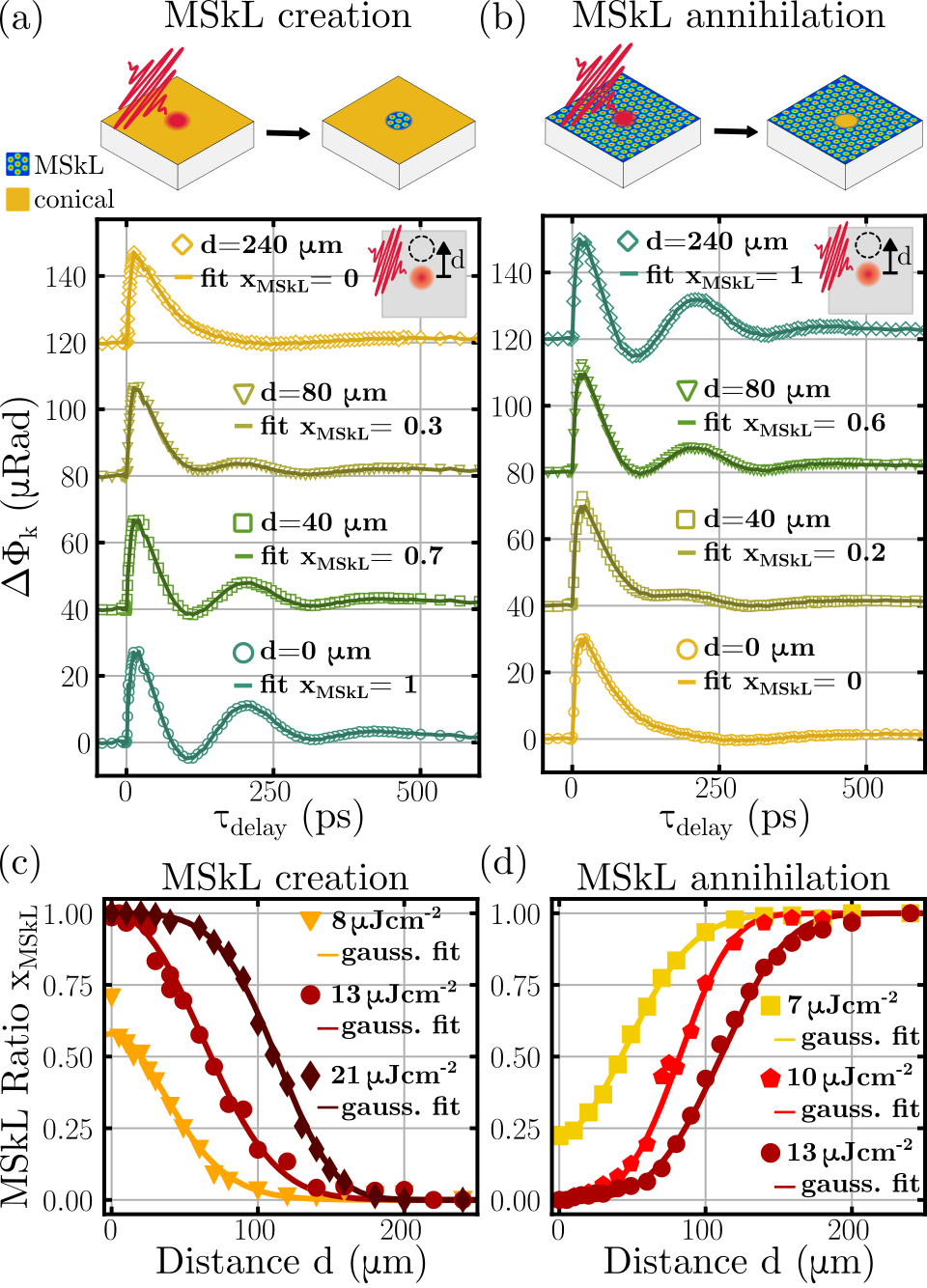}
\caption{\label{fig:Figure2} Spatially resolved magnetization dynamics of Fe$_{0.75}$Co$_{0.25}$Si after laser-induced MSkL creation (${55\,\text{mT}}$) and annihilation (${70\,\text{mT}}$) at ${13\,\text{K}}$: (a) and (b) Magnetization dynamics measured by TR-MOKE at different distances $d$ from the irradiation spot. The solid lines represent fits to the experimental data using Eq. \ref{eq:superposition}. (c) and (d) MSkL ratio ${x_{\text{MSkL}}}$ in the probed area as a function of distance to the irradiation spot for laser-induced MSkL creation and annihilation after irradiation with different laser fluences.  The solid lines represent fits to the experimental data using the model introduced in the main text.
}
\end{figure}

Complementing the observation of a MSkL after homogeneous rapid field cooling of cubic chiral magnets in previous studies, a local heating process was implemented in terms of a series of high-fluence fs laser pulses in this study. The cumulative heat load of the laser pulses leads to a local increase of the sample temperature in a spatially confined volume in the vicinity of the laser focal spot, which is much smaller than the sample dimension of ${(2\times2\times2)\,\text{mm}^3}$, see thermal simulations in Ref. \cite{supplement}. Experimentally, this condition is realized by increasing the laser fluence of the TR-MOKE pump beam for less than a second. After reducing the laser fluence again, the local temperature increase is rapidly compensated by the thermal reservoir provided by the cryostat resulting in a fast and local cryogenic sample cooling. As a first implementation test of this local heating process, we start in the conical state at ${13\,\text{K}}$ and ${55\,\text{mT}}$, see circle in Fig. \ref{fig:Figure1}(a), and apply high-fluence laser pulses of ${F=13\,\mu\text{Jcm}^{-2}}$ to the sample.
After the high-fluence laser irradiation and subsequent cryogenic cooling, the magnetic state is probed spatially-resolved in TR-MOKE experiments. Figure \ref{fig:Figure2}(a) exemplarily shows magnetization dynamics at four sample positions with different distances $d$ to the center of the laser irradiation spot. While for large distances, ${d=240\mu\text{m}}$, we observe the characteristic dynamics of the initial conical state, in proximity to the irradiation spot the precessional dynamics of the skyrmion breathing mode emerge. With increasing distance, the magnetization dynamics gradually evolve from the characteristic behavior of the MSkL to that of the conical state. 
This result shows that local laser irradiation of the sample can drive a conical-to-MSkL transition in a confined area and thus permits the laser-induced creation of a skyrmion patch in a bulk chiral magnet. By that, we are able to locally create a thermodynamically metastable state with non-trivial topology, which persists over long time scales (${> 12\,\text{h}}$, not shown) and is embedded in a topologically trivial equilibrium phase. This laser-written state shows a comparable magnetic field and temperature stability as the homogeneous MSkL created by rapid field cooling \cite{supplement}, demonstrating the high robustness of the spatially confined MSkL. As we observe the laser-induced magnetic phase transition only for magnetic fields in the regime of the SkL \cite{supplement}, we attribute the formation of the laser-written MSkL to local thermal quenching of the equilibrium SkL by laser heating.

Using a similar approach, the inverse process can be achieved: laser-induced MSkL annihilation. Following the formation of a spatially homogeneous MSkL by means of rapid field cooling to ${13\,\text{K}}$, the magnetization dynamics are studied after irradiation with high-fluence laser pulses and subsequent cryogenic cooling at ${70\,\text{mT}}$, i.e., at a magnetic field above the SkL pocket, see circle in Fig. \ref{fig:Figure1}(b). In a situation reverse to laser-induced MSkL creation, we observe a local formation of the conical state in proximity to the irradiation spot, as shown in Fig. \ref{fig:Figure2}(b). 
Dynamics characteristic of the initial MSkL is detected far from the laser irradiation spot, continuously evolving into the dynamics of the conical state with decreasing distance. As anticipated from the phase diagram in Fig. \ref{fig:Figure1}(b), a switching from MSkL to the conical state after laser irradiation occurs exclusively for magnetic field values outside the equilibrium SkL \cite{supplement}. Consequently, we attribute the laser-induced annihilation of the MSkL to its thermal destabilization by laser heating. In line with our understanding of MSkL creation in the field regime of the equilibrium SkL, the MSkL persists after intense laser irradiation within this regime
Thus, laser-induced MSkL creation and annihilation takes place in well-defined and separated magnetic field regimes, allowing selective optical writing and deleting of MSkL controlled by the applied magnetic field. From an application point of view, this might enable all-optical MSkL creation and annihilation sequences with readout operation without the need of rapid temperature control and extensive magnetic field sweeps.

The results presented so far demonstrate that laser-induced MSkL creation and annihilation leads to a gradual phase transition in space.
The spatially resolved measurements can be modeled by a linear superposition of the conical, $\Delta \phi_{k,\text{conical}}$, and MSkL contributions, $\Delta \phi_{k,\text{MSkL}}$, to the measured signal
\begin{equation}
\Delta \phi_k(\tau_{\text{delay}}) = x_{\text{MSkL}} \cdot \Delta \phi_{k,\text{MSkL}}(\tau_{\text{delay}})  + \left(1-x_{\text{MSkL}}\right)\cdot \Delta \phi_{k,\text{conical}}(\tau_{\text{delay}}). \label{eq:superposition}
\end{equation}
The MSkL ratio ${x_{\text{MSkL}}}$ describes the areal percentage of MSkL in the probed region, as the TR-MOKE technique averages over magnetization dynamics in the probed area. The line plots in Figs. \ref{fig:Figure2}(a) and \ref{fig:Figure2}(b) are fits to the experimental data using Eq. \ref{eq:superposition} with ${x_{\text{MSkL}}}$ as fitting parameter, in excellent agreement with the data. By plotting ${x_{\text{MSkL}}}$ as a function of distance $d$ from the laser irradiation spot, as shown in Fig. \ref{fig:Figure2} (c) and \ref{fig:Figure2}(d) after laser irradiation with different fluences $F$, 
the spatial dimensions of the patches of laser-induced MSkL creation and annihilation may be determined quantitatively. Extracting the spatial MSkL distribution ${\rho_{\text{MSkL}}(r)}$ requires to deconvolve ${x_{\text{MSkL}}(r)}$ with the intensity profile of the probe beam ${\frac{I(r+d)}{I_0}}$ \cite{supplement} to take into account the spatial averaging of TR-MOKE.
The spatial MSkL distribution after laser-induced MSkL creation is best described by a high-order Gaussian function given by
\begin{equation}
\rho_{\text{MSkL}}(r) =  \rho_0 \cdot  \exp\left( \frac{-2|r|^{n}}{|w_\text{MSkL}|^{n}} \right), \label{eq:SuperGaussianDistribution}
\end{equation}
with radius ${w_{\text{MSkL}}}$ corresponding to the distance from the center where ${\rho_{\text{MSkL}}}$ drops to $1/e^2$.
For laser-induced MSkL annihilation, the MSkL distribution is described by ${1-\rho_{\text{MSkL}}(r)}$.
The line plots in Figs. \ref{fig:Figure2}(c) and \ref{fig:Figure2}(d) show that the experimental data can be well approximated with the chosen model using ${w_{\text{MSkL}}}$, $n$, and $\rho_0$ as fit parameters. The resulting values are listed in Ref. \cite{supplement}.
The radius ${w_{\text{MSkL}}}$ increases with the laser fluence and gets larger than the laser focal width for high fluences, consistent with sizable thermal diffusion, see thermal simulations in Ref. \cite{supplement}. Moreover, the shape of the MSkL distribution changes with laser fluence. Irradiation with smaller laser fluence results in the formation of a Gaussian MSkL distribution ($n\leq2$), shown for ${8\,\mu\text{Jcm}^{-2}}$ and ${7\,\mu\text{Jcm}^{-2}}$ for laser-induced MSkL creation and annihilation in Figs. \ref{fig:Figure2}(c) and \ref{fig:Figure2}(d). For higher laser fluence, the MSkL distribution forms a flat top ($n>2$) that broadens under further increasing laser fluences. The spatial MSkL distribution ${\rho_{\text{MSkL}}}$ differs for laser-induced creation and annihilation as reflected by different ${w_{\text{MSkL}}}$ and $n$. Most notably, for the same irradiation fluence of ${13\,\mu\text{Jcm}^{-2}}$, the MSkL is annihilated in a larger region than it is created, leading to different patch sizes of laser-induced MSkL creation and annihilation.

\begin{figure}
\includegraphics{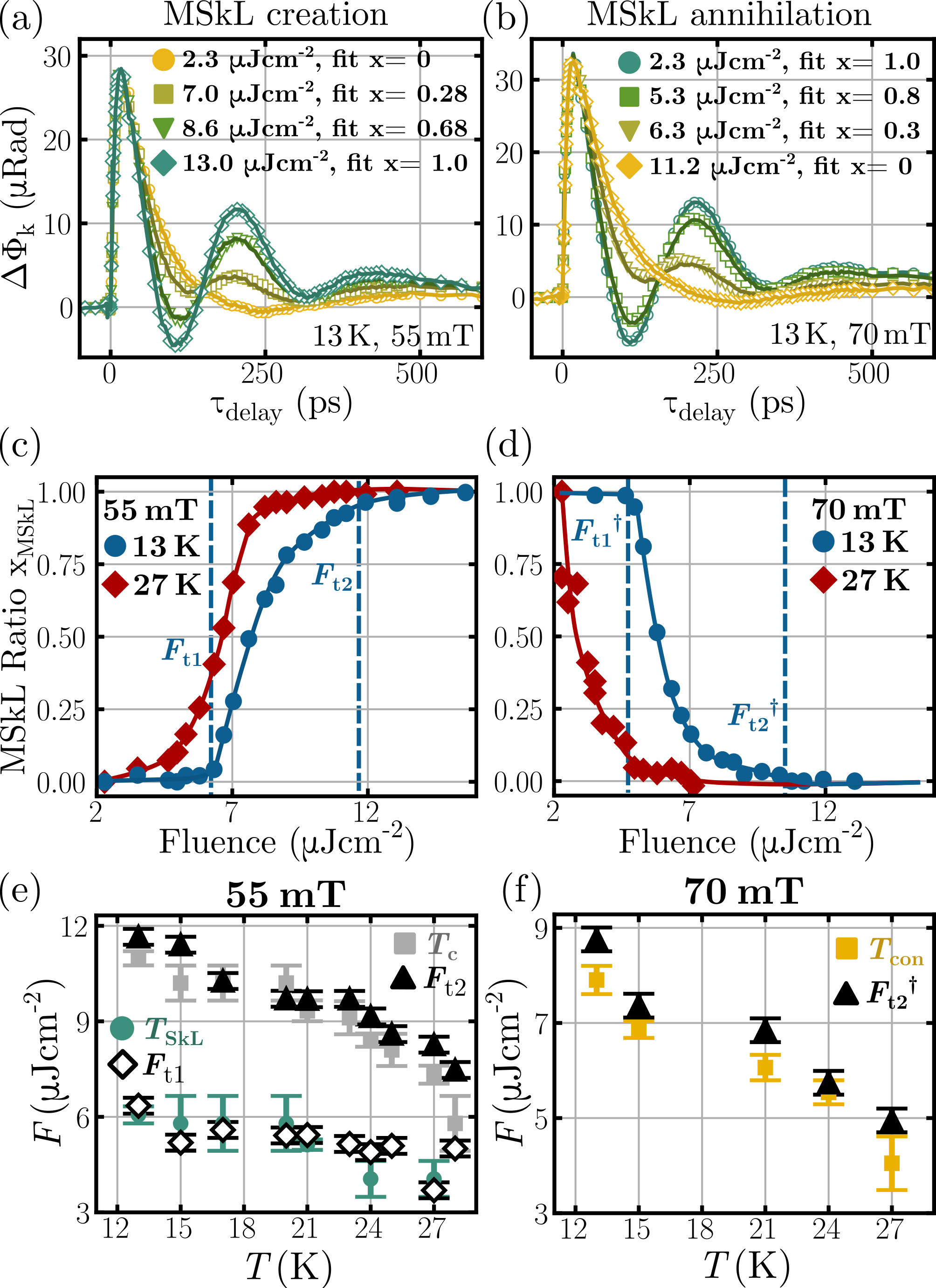}
\caption{\label{fig:Figure3} Fluence dependence of laser-induced MSkL creation and annihilation in Fe$_{0.75}$Co$_{0.25}$Si: (a) and (b) magnetization dynamics after the irradiation with fs laser pulses of different laser fluences starting in the equilibrium conical state (a) and MSkL (b). (c) and (d) MSkL ratio ${x_{\text{MSkL}}}$ as a function of irradiation fluence for laser-induced MSkL creation (c) and annihilation (d) at ${13\,\text{K}}$ and ${27\,\text{K}}$. The dashed blue lines show exemplary the 5\% and 95\% fluence thresholds at ${13\,\text{K}}$. (e) and (f) Comparison of the laser fluence required to increase the sample temperature to $T_{\text{c}}$, $T_{\text{SkL}}$ and $T_{\text{con}}$ for different temperatures and the fluence thresholds of laser-induced MSkL creation ($F_{\text{t1}}$, $F_{\text{t2}}$) and annihilation ($F_{\text{t2}}^\dagger$). 
}
\end{figure}
This finding implies the need to investigate the fluence dependence of these process, which we accomplish by means of analyzing the magnetic state at the center of the laser spot (${d=0\,\mu\text{m}}$). Figures \ref{fig:Figure3}(a) and \ref{fig:Figure3}(b) show TR-MOKE measurements after laser irradiation with four different laser fluences and subsequent cryogenic cooling starting in the equilibrium conical state (${55\,\text{mT}}$, ${13\,\text{K}}$) and the laser-written MSkL (${70\,\text{mT}}$, ${13\,\text{K}}$), respectively. With increasing fluence, the magnetization dynamics gradually evolves from the characteristic behavior of the conical state to that of the MSkL and vice versa. Again, the fluence-dependent measurements are described well by Eq. \ref{eq:superposition}, see solid lines in Figs. \ref{fig:Figure3}(a) and \ref{fig:Figure3}(b). In Figs. \ref{fig:Figure3}(c) and \ref{fig:Figure3}(d) for two different temperatures the evolution of the MSkL ratio ${x_{\text{MSkL}}}$ is presented as a function of laser fluence for MSkL creation and annihilation, respectively. 
We observe a double-threshold behavior for laser-induced MSkL creation ($F_{\text{t1}}$, $F_{\text{t2}}$) and annihilation ($F_{\text{t1}}^{\dagger}$, $F_{\text{t2}}^{\dagger}$). The laser-induced magnetic phase transition sets in for fluences larger than a lower threshold $F_{\text{t1}}^{(\dagger)}$ and completes above an upper fluence threshold $F_{\text{t2}}^{(\dagger)}$.  
In the fluence regime between $F_{\text{t1}}^{(\dagger)}$ and $F_{\text{t2}}^{(\dagger)}$ the MSkL ratio ${x_{\text{MSkL}}}$ scales with the irradiation fluence.
Thus, for fluences smaller than $F_{\text{t2}}^{(\dagger)}$, the laser irradiation leads to only a partial switching between MSkL and equilibrium conical state and to a coexistence of both states with a fluence-dependent MSkL ratio.

In Figs. \ref{fig:Figure3}(c) and \ref{fig:Figure3}(d) the fluence thresholds $F_{\text{t1}}^{(\dagger)}$ and $F_{\text{t2}}^{(\dagger)}$, approximated by 5\% and 95\% switching ratios toward the final state, are shown as dashed vertical lines. The comparison of the thresholds for laser-induced MSkL creation and annihilation shows that the annihilation of the MSkL sets in and completes at lower fluences than the creation process at the same temperature. With that, the different patch sizes of laser-induced MSkL creation and annihilation can be traced back to the different fluence dependencies. Namely, by irradiation with the same laser fluence the lower fluence threshold for MSkL annihilation is exceeded in a larger area than the higher threshold for MSkL creation. 
Moreover, the transition from a Gaussian to a flat-top MSkL distribution, characterized by an increase of $n$ in Eq. \ref{eq:SuperGaussianDistribution}, aligns closely with the concept of laser-induced MSkL creation and annihilation as a two-threshold process. This is evident by the saturation of the final state at the center of the irradiation beam when locally reaching the threshold $F_{\text{t2}}^{(\dagger)}$.

Figs. \ref{fig:Figure3}(c) and \ref{fig:Figure3}(d) demonstrate that the fluence thresholds of laser-induced MSkL creation and annihilation 
decreases at higher temperatures, consistent with the laser-induced temperature rise in the material being key for MSkL formation. To quantitatively assess this, we estimate the temperature increase caused by laser heating by comparing fluence and temperature-dependent TR-MOKE measurements \cite{supplement}. In Fig. \ref{fig:Figure3}(e), the laser fluence values leading to a local increase of the sample temperature to $T_{\text{SkL}}$ (turquoise circles) and $T_{\text{c}}$ (gray square) are compared to the upper $F_{\text{t1}}$ (white diamonds) and lower $F_{\text{t2}}$ (black triangle) fluence threshold of laser-induced MSkL creation. Here, $T_{\text{SkL}}$ represents the lowest temperature within the SkL regime for a given magnetic field, as illustrated in Fig. \ref{fig:Figure1}(a).
Notably, Fig. \ref{fig:Figure3}(e) reveals that $F_{\text{t1}}$ and $F_{\text{t2}}$ correspond to a local temperature increase to $T_{\text{SkL}}$ and $T_{\text{c}}$, respectively. This aligns well with laser-induced MSkL creation due to local thermal quenching, as the equilibrium SkL stabilizes in this temperature regime and is then subsequently quenched. The increase of the MSkL ratio in the fluence regime from $F_{\text{t1}}$ to $F_{\text{t2}}$, i.e., for higher temperature rises in the SkL, might be attributed to the speedup of the topological winding process for temperatures approaching $T_{\text{c}}$. This influences the probability of SkL stabilization and consequently affects the number of skyrmions preserved following laser heating. In agreement with the phase diagram shown the Fig. \ref{fig:Figure1}b) local temperature increases to $T_{\text{con}}$ corresponds to the upper threshold of MSkL annihilation $F_{\text{t2}}^\dagger$, as shown in Fig. \ref{fig:Figure3}(f). 
This confirms that laser-induced MSkL annihilation can be explained by local transient laser heating of the sample to temperatures that destabilize the MSkL and lead to the formation of the conical state.
Additionally, this study demonstrates that even temperature rises much smaller than $T_{\text{con}}$ are sufficient to partially switch from MSkL to the conical state, likely due to the temperature-dependent lifetime of the MSkL. As temperature increases the MSkL lifetime, representing the characteristic time scale for skyrmion unwinding, decreases, resulting in an increased transition probability between the MSkL and conical state during high fluence laser irradiation.

In summary, our study explores the all-optical manipulation of a spatially-confined metastable skyrmion lattice state (MSkL) in a bulk compound of the cubic chiral magnet Fe$_{0.25}$Co$_{0.75}$Si. 
Probed by the magnetization dynamics in TR-MOKE experiments, we demonstrate the feasibility of laser-induced MSkL creation and annihilation in spatially confined areas for well-defined magnetic field regimes. 
Spatially resolved and fluence-dependent measurements indicate that creation and annihilation in Fe$_{0.25}$Co$_{0.75}$Si are linked to 
distinct temperature regimes for the stabilization and destabilization of the MSkL.
Aside from promising steps towards the all-optical local control of skyrmions for a fast and energy efficient manipulation in future storage or microwave devices \cite{wang2020ultrafast}, our study provides insights into optically-induced topological phase transitions in terms of thermal quenching.

\begin{acknowledgments}
\textbf{Funding:} This work was supported by the European Metrology Research Programme (EMRP)
and EMRP participating countries under the European Metrology Programme for Innovation
and Research (EMPIR) Project No. 17FUN08-TOPS Metrology for topological spin structures. In part, this study has been funded by the Deutsche Forschungsgemeinschaft (DFG, German ResearchFoundation) under TRR80 (From Electronic Correlations to Functionality, Project No.\ 107745057, Project E1), SPP2137 (Skyrmionics, Project No.\ 403191981, Grant PF393/19 and Grant SCHU 2250/8-1), the excellence cluster MCQST under Germany's Excellence Strategy EXC-2111 (ProjectNo.\ 390814868) and EXC-2123 QuantumFrontiers (ProjectNo. 390837967). Financial support by the European Research Council (ERC) through Advanced Grants No.\ 291079 (TOPFIT) and No.\ 788031 (ExQuiSid) is gratefully acknowledged.
\textbf{Author contributions:} JK initiated the study. AB and CP grew the crystal. JK performed the TR-MOKE study under the supervision of MB and HF. The data analysis was carried out by JK. DK and RA supported the study with simulations. MB, HF, SS, HWS, DK, RA, JK, AB and CP discussed the results. JK wrote the paper.
\textbf{Competing interests:} The authors declare that they have no competing interests. 
\textbf{Data availability}: All data needed to evaluate the conclusions of the paper are present in the paper and in the supplemental material. 
\end{acknowledgments}

\nocite{*}
\bibliography{References}

\end{document}



\title[Supplemental Material: Optical creation and annihilation of skyrmion patches in a bulk chiral magnet]{Supplemental Material: Optical creation and annihilation of skyrmion patches in a bulk chiral magnet}
\author{J. Kalin$^*$}\email{jantje.kalin@ptb.de}
\affiliation{Physikalisch-Technische Bundesanstalt, 38116 Braunschweig, Germany}

\author{D. Kalin}
\affiliation{ESE Engineering und Software-Entwicklung GmbH, 38122 Braunschweig, Germany}

\author{S. Sievers}%
\affiliation{Physikalisch-Technische Bundesanstalt, 38116 Braunschweig, Germany}

\author{A. Bauer}
\affiliation{Physik-Department, Technische Universität München, 85748 Garching, Germany}
\affiliation{Zentrum für QuantumEngineering (ZQE), Technische Universität München, 85748 Garching, Germany}

\author{H. W. Schumacher}
\affiliation{Physikalisch-Technische Bundesanstalt, 38116 Braunschweig, Germany}

\author{R. Abram}
\affiliation{Physikalisch-Technische Bundesanstalt, 38116 Braunschweig, Germany}

\author{H. Füser}
\affiliation{Physikalisch-Technische Bundesanstalt, 38116 Braunschweig, Germany}

\author{C. Pfleiderer}
\affiliation{Physik-Department, Technische Universität München, 85748 Garching, Germany}
\affiliation{Zentrum für QuantumEngineering (ZQE), Technische Universität München, 85748 Garching, Germany}
\affiliation{Munich Center for Quantum Science and Technology (MCQST), Technische Universität München, 85748 Garching, Germany}

\author{M. Bieler}
\affiliation{Physikalisch-Technische Bundesanstalt, 38116 Braunschweig, Germany}

\date{\today}

\maketitle

The Supplemental Material contains further information on (I) the magnetic field dependence of laser-induced MSkL creation and annihilation, (II) the temperature and magnetic field stability of the laser-written MSkL state, (III) the fluence dependence of laser-induced MSkL creation and annihilation, (IV) spatially-resolved measurements and modeling, (V) estimation of laser heating by the irradiation with femtosecond laser pulses and (VI) finite element modeling of laser heating in Fe$_{0.75}$Co$_{0.25}$Si.

\section{Magnetic field dependence of laser-induced MSkL creation and annihilation}
\begin{figure}[!]
\includegraphics{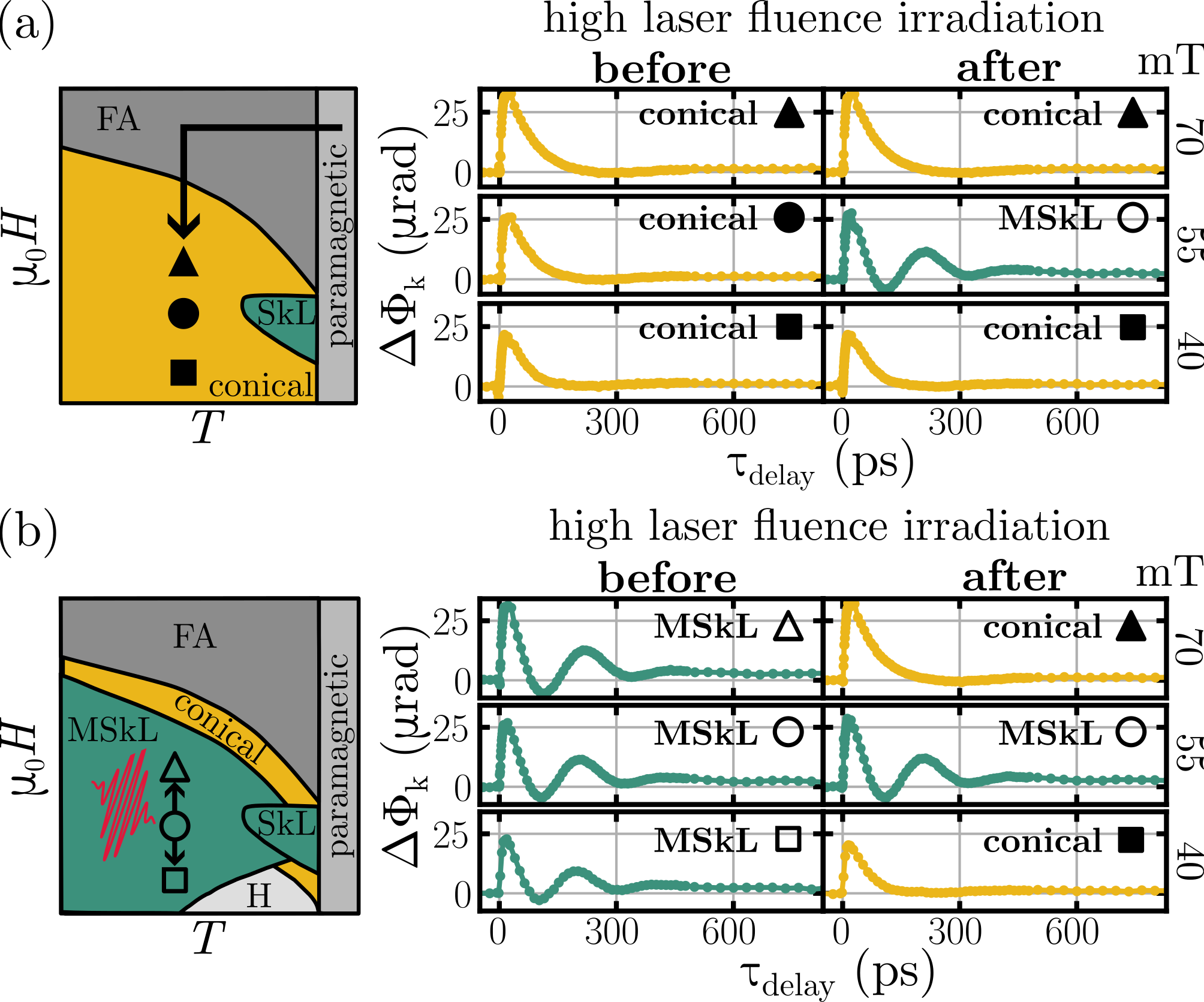}
\caption{\label{fig:FigureS1} Magnetization dynamics of Fe$_{0.75}$Co$_{0.25}$Si at $15\,$K before and after high laser fluence irradiation and subsequent cryogenic cooling at $40\,$mT ($\square$, lower panel), $55\,$mT ($\iscircle$, center panel) and $70\,$mT ($\triangle$, upper panel) starting in the equilibrium conical state (a) and MSkL state (b).
}
\end{figure}

In this section, we present further details about the magnetic field dependence of the laser-induced MSkL creation and annihilation process.
To that end, we prepare the conical state (a) and MSkL state (b) at three characteristic magnetic field values: $H = 40\,$mT ($\square$, below SkL), $H = 55\,$mT ($\iscircle$, center SkL), and $H = 70\,$mT ($\triangle$, above SkL). 
The appearance of these states for all field values is confirmed by TR-MOKE measurements revealing the characteristic out-of-plane magnetization dynamics shown in the center panel of Fig. \ref{fig:FigureS1}(a) and (b). Subsequently, we apply a series of high-fluence laser pulses ($F=13\,\mu$Jcm$^{-2}$) to the sample and let the sample cooldown rapidly by the thermal bath provided by the cryostat after we reduce the laser fluence. The right panel of  Fig. \ref{fig:FigureS1} (a) and (b) shows the magnetization dynamics observed after laser irradiation of the sample and subsequent cryogenic cooling.
Only for magnetic field values in the field span of the equilibrium SkL, here shown for $55\,$mT, we observe in Fig. \ref{fig:FigureS1}(a) laser-induced MSkL creation, i.e., the transition from the conical to MSkL state. In comparison, the phase transition from the MSkL to the conical state, which results from laser-induced MSkL annihilation, takes only place for magnetic field values outside the field span of the equilibrium SkL, see right panel of Fig. \ref{fig:FigureS1}(b).

\section{Temperature and magnetic field stability of the laser-written MSkL state}

\begin{figure}[!]
\includegraphics{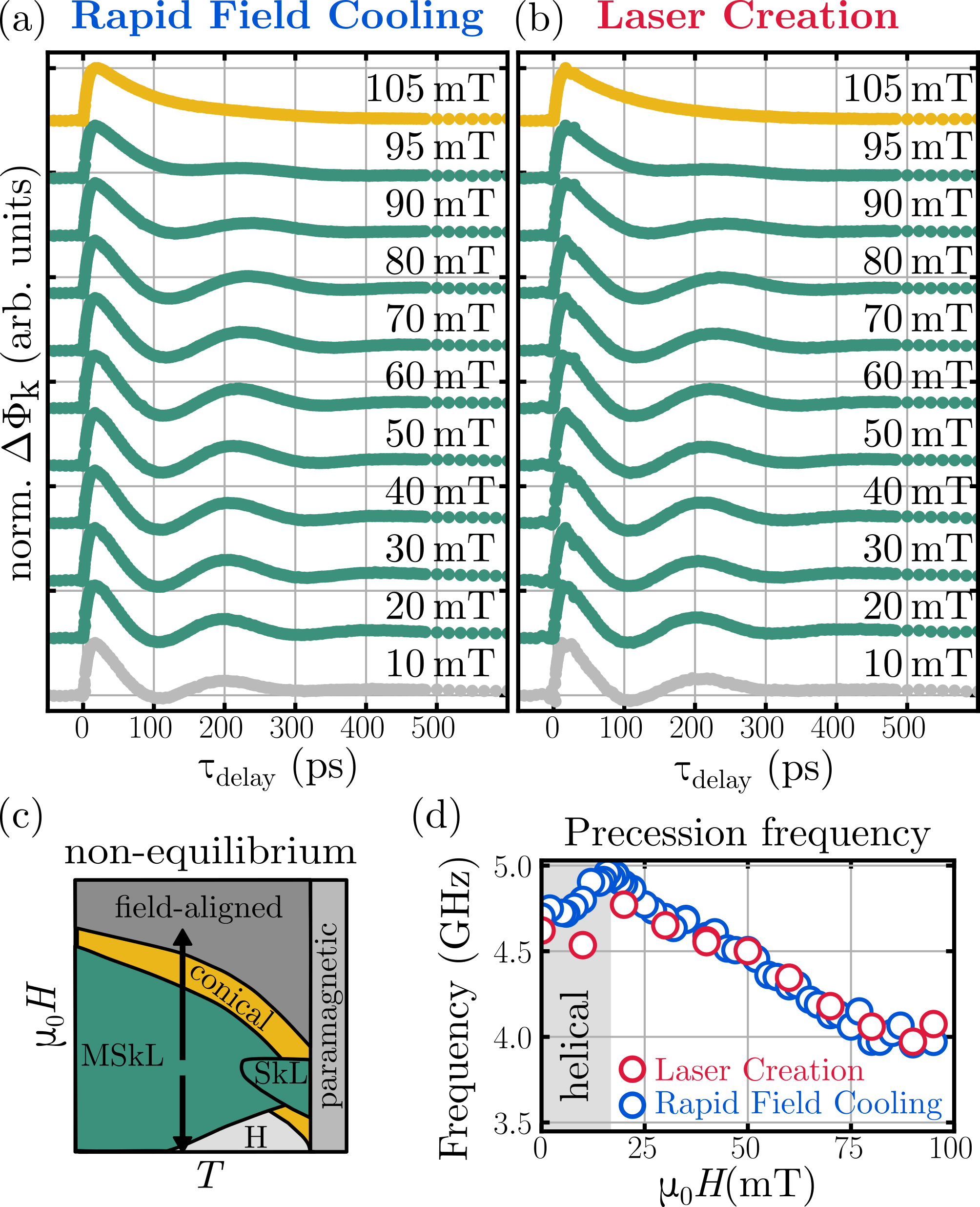}
\caption{\label{fig:FigureS2} Magnetization dynamics of Fe$_{0.75}$Co$_{0.25}$Si at $15\,$K for different magnetic fields after (a) rapid field cooling and (b) laser-induced MSkL creation at $55\,$mT. The measurements were collected in field sweeps with increasing and decreasing fields starting at $55\,$mT as shown in (c). (d) Precession frequency as a function of the applied magnetic field for rapid field cooling (blue circle) and laser-induced MSkL creation (red circle).
}
\end{figure}

By temperature and magnetic field dependent TR-MOKE measurements we characterize the magnetization dynamics of the laser-written MSkL state and compare it to the results achieved after thermal quenching by rapid field cooling. 
Figure \ref{fig:FigureS2} shows the magnetic field dependent magnetization dynamics of the MSkL state at $15\,$K prepared by rapid field cooling (a) and irradiation with high-F laser pulses (b). In rapid field cooling the sample is cooled down at $55\,$mT from $T=60\,$K$>T_{\text{c}}$ to $15\,$K by cryogenic temperature control. Laser-induced MSkL creation is performed at the same magnetic field value at $15\,$K. The magnetization dynamics measured by TR-MOKE are recorded in field sweeps for increasing and decreasing magnetic fields (see Fig. \ref{fig:FigureS2}(c)). Before each field sweep the MSkL creation procedure is repeated. 
By analyzing the magnetization dynamics with the phenomenological model proposed in Ref. [9], we determine the precession frequency against magnetic field, see Fig. \ref{fig:FigureS2}(d).
The laser-written MSkL state persists in the same magnetic field range ($20-95\,$mT) than the one prepared by rapid field cooling, indicating a comparable stability against magnetic field changes. For magnetic fields above $95\,$mT the MSkL state vanishes by the transition to the equilibrium conical state, which shows a non-precessional de- and remagnetization process. Note here, that the phase transition from the MSkL to the conical state is of second order and takes place via the formation of a phase coexistence. 
The decrease of the MSkL precession frequency with increasing magnetic field, shown in Fig. \ref{fig:FigureS2}(d), identifies the observed dynamics as the skyrmion breathing mode. The change of frequency with magnetic field is thereby the same for the MSkL state prepared by laser irradiation and rapid field cooling.

\begin{figure}[!]
\includegraphics{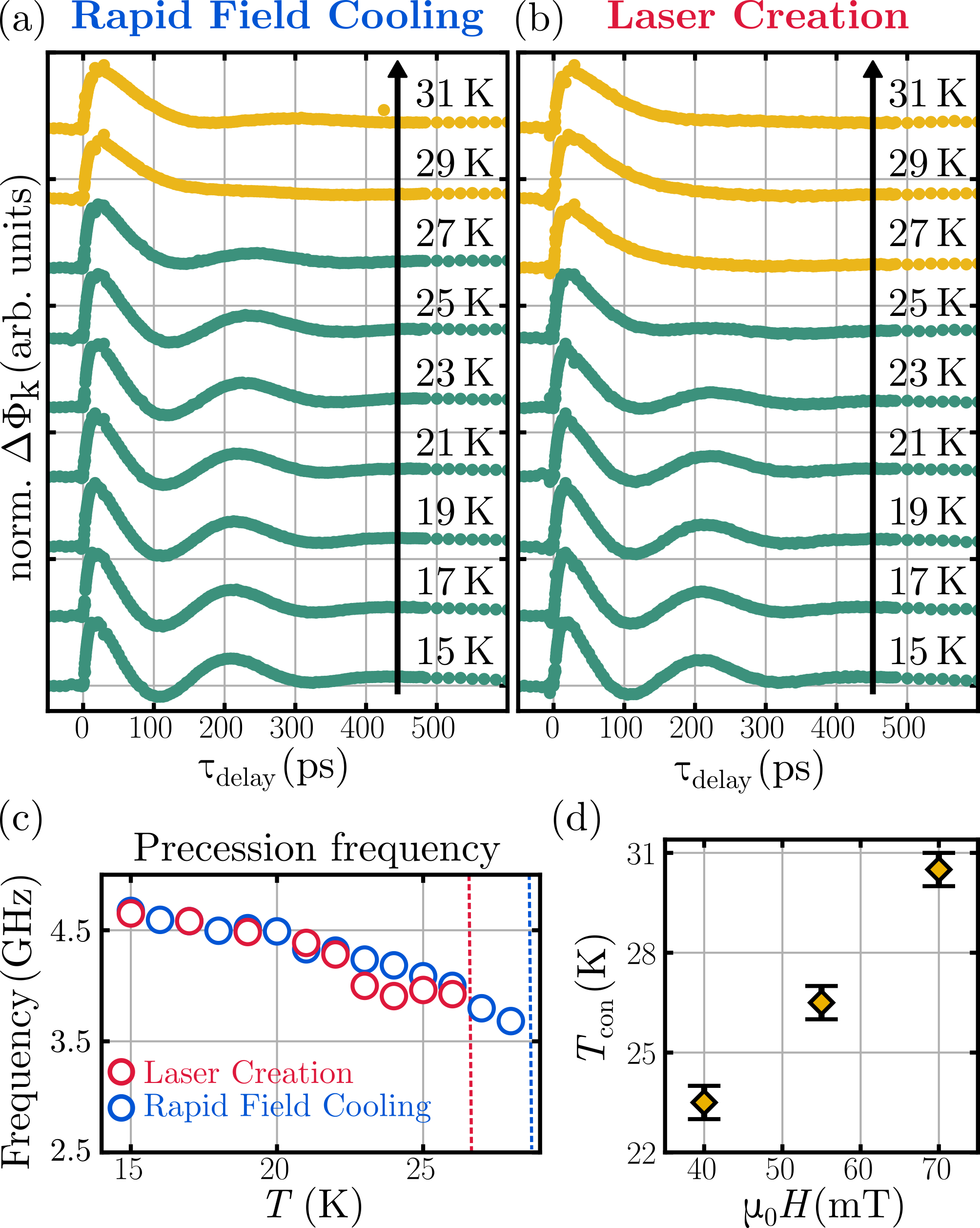}
\caption{\label{fig:FigureS3} (a) Magnetization dynamics of the MSkL state in Fe$_{0.75}$Co$_{0.25}$Si at $55\,$mT for different temperatures created by (a) rapid field cooling to $15\,$K at $55\,$mT and (b) by laser pulse irradiation at $15\,$K and $55\,$mT. (c) Precession frequency as a function of the temperature after rapid field cooling (blue circle) and laser-induced MSkL creation (red circle). The dashed lines show the transition temperature  $T_{\text{con}}$ from MSkL to equilibrium conical state for laser-induced MSkL creation (red line) and rapid field cooling (blue line). (d) $T_{\text{con}}$ for the laser-written MSkL state as a function of the applied magnetic field.
}
\end{figure}

Next, we study in Fig. \ref{fig:FigureS3} the temperature dependent dynamics of the laser-written MSkL state (b) and compare it again to the MSkL state prepared by rapid field cooling (a). To that end, we create the MSkL state at $15\,$K and $55\,$mT and characterize the magnetization dynamics after temperature increase. We extract the precession frequency again by a fit to the phenomenological model from Ref. [9] and plot it in Fig. \ref{fig:FigureS3}(c).
The dynamics of the laser-written MSkL state shows a comparable temperature dependence than the one prepared by rapid field cooling. The precession frequency decreases for higher temperatures and the MSkL state vanishes upon transition to the equilibrium conical state at the transition temperature $T_{\text{con}}$.  For the laser-written MSkL state the transition occurs for slightly lower temperatures ($T_{\text{con}}$=$26\,$K) than for the MSkL state prepared by rapid field cooling ($T_{\text{con}}$=$28\,$K). Again, we observe a second order phase transition and thus a formation of a phase coexistence of the MSkL and conical state for increasing temperature. As shown in Fig. \ref{fig:FigureS3}(d) the transition temperature $T_{\text{con}}$ increases with increasing magnetic field. This indicates that the MSkL state possesses a higher temperature stability at higher magnetic fields. 

\section{Fluence dependence of laser-induced MSkL creation and annihilation}
\begin{figure}[!]
\includegraphics{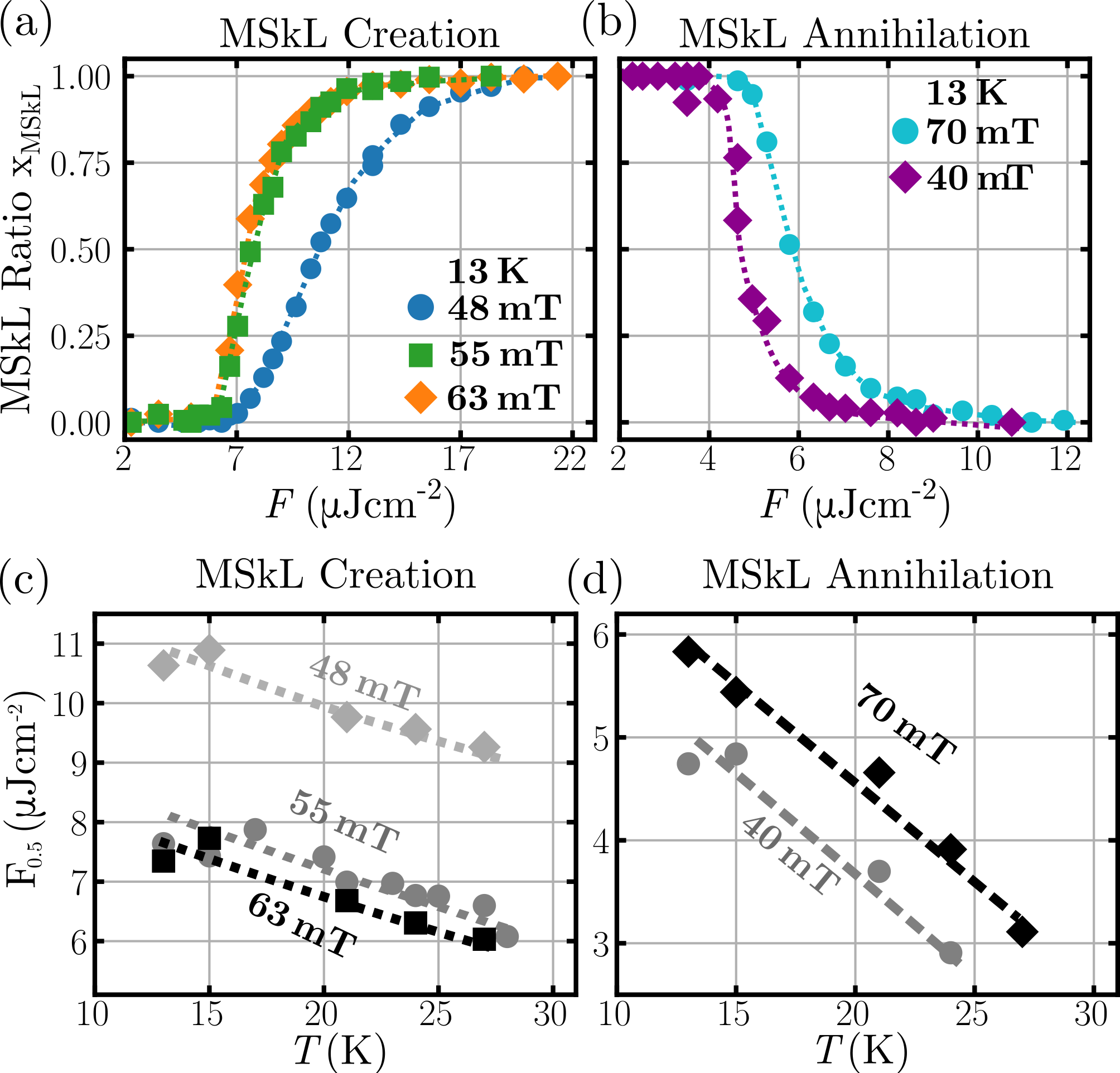}
\caption{\label{fig:FigureS4} Fluence dependence of laser-induced MSkL creation and annihilation in Fe$_{0.75}$Co$_{0.25}$Si: MSkL ratio $x_{\text{MSkL}}$ as a function of irradiation fluence at $13\,$K for (a) laser-induced MSkL creation at $48\,$mT (blue circle), $55\,$mT (green square) and $63\,$mT (orange diamond) and for (b) laser-induced MSkL annihilation at $40\,$mT (purple diamonds) and $70\,$mT (light blue circle). Characteristic threshold value $F_{0.5}$ as a function of temperature for (c) laser-induced MSkL creation at $48\,$mT (gray diamond), $55\,$mT (dark gray circle) and $63\,$mT (black square) and (d) laser-induced MSkL annihilation at $40\,$mT (dark gray circle) and $70\,$mT (black diamond).
}
\end{figure}

Here, we give further information about the fluence dependence of laser-induced MSkL creation and annihilation in Fe$_{0.25}$Co$_{0.75}$Si.
In Fig. \ref{fig:FigureS4} (a) and (b) we show the fluence-dependent MSkL ratio for laser-induced MSkL creation and annihilation  at $13\,$K for different magnetic fields.
For laser-induced MSkL creation, we observe that the lower fluence threshold $F_{t1}$, where the process sets in, only changes slightly for different magnetic field values. However, the upper fluence threshold $F_{t2}$ and the fluence value of $50$\% MSkL creation $F_{0.5}$ strongly increase for lower magnetic fields, shown exemplary for $48\,$mT. 
In contrast, for laser-induced MSkL annihilation, all characteristic threshold values ($F_{t1}^\dagger$, $F_{t2}^\dagger$, $F_{0.5}$) decrease similarly with increasing magnetic field. Consequently we need less laser fluence to annihilate MSkL at lower magnetic fields.
The temperature dependence of $F_{0.5}$ for laser-induced MSkL creation and annihilation is shown for the different magnetic field values as a function of temperature in Fig. \ref{fig:FigureS4}(c) and (d), respectively. Overall we observe a decrease of $F_{0.5}$ with increasing temperature, however, with different slopes for laser-induced MSkL creation and annihilation. 

\section{Spatially resolved measurements and modeling}
\begin{figure}
\includegraphics{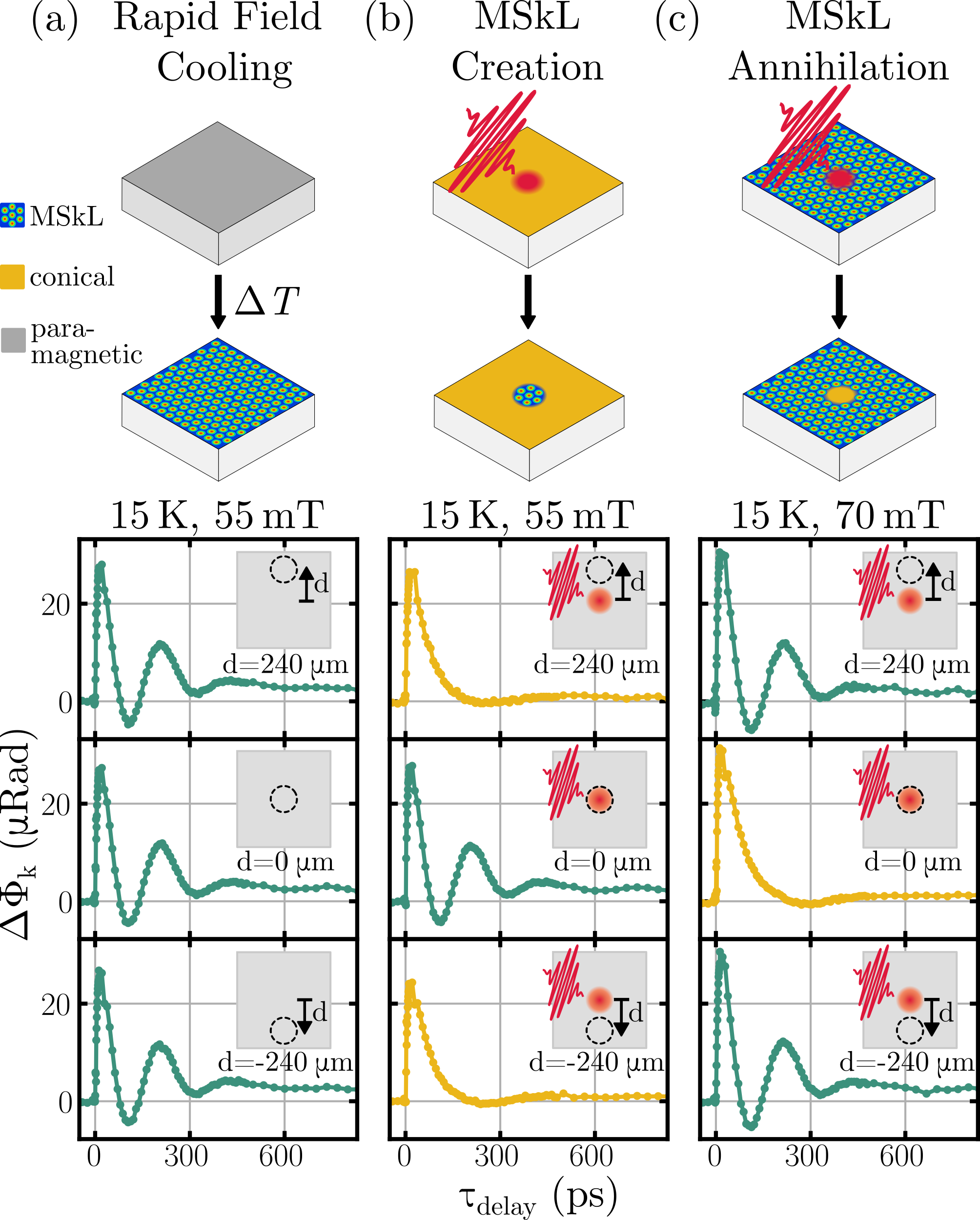}
\caption{\label{fig:FigureS5} Magnetization dynamics of Fe$_{0.75}$Co$_{0.25}$Si at three characteristic sample positions, $d=0\,\mu$m at the center of the laser irradiation spot and $d=\pm240\,\mu$m far from the laser irradiation spot, are shown (a) after rapid field cooling ($55\,$mT, $13\,$K), (b) after laser-induced creation of MSkL ($55\,$mT, $13\,$K) and (c) after laser annihilation of MSkL ($70\,$mT, $13\,$K).
}
\end{figure}

\begin{figure}
\includegraphics{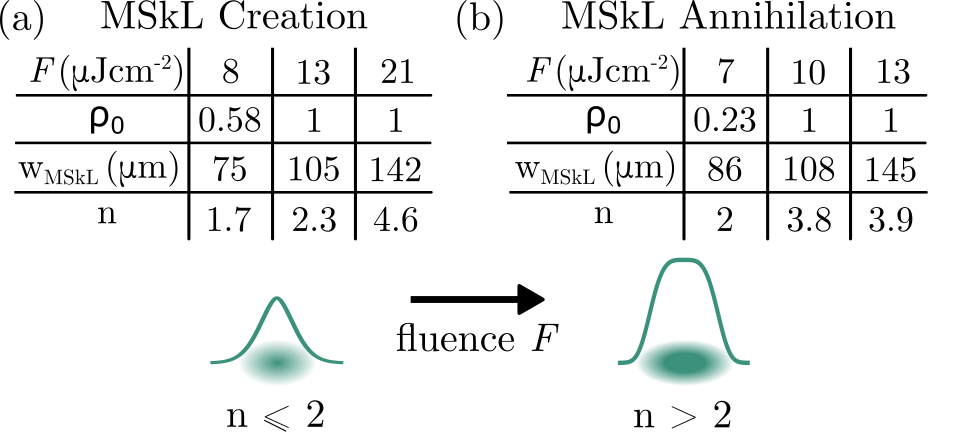}
\caption{\label{fig:FigureS6} Modeling of MSkL distribution: Fit parameters of the line plots in Fig. 2(c) and (d) in the main text. 
}
\end{figure}

In this section, we provide further information about the spatially resolved TR-MOKE measurements and the corresponding modeling. 
We start with a comparison of the spatial expansion of the MSkL state created by rapid field cooling and laser-induced creation and annihilation.
Figure \ref{fig:FigureS5} shows the magnetization dynamics measured at three characteristic sample positions: in the center of the laser irradiation spot $d=0\,\mu$m and far from the irradiation spot $\pm240\,\mu$m. As described in the main text the laser-induced MSkL creation and annihilation is a spatially confined process. In comparison, rapid field cooling leads to a formation of a homogeneous MSkL state all over the sample. Remarkably, even the precession amplitude is the same for the three shown sample positions, indicating a homogeneous MSkL density all over the sample. 
In order to analyze the spatially resolved measurements we set up a model to quantitatively describe the laser-induced MSkL distribution. In the model two experimental aspects are taken into account. First, the laser-induced MSkL creation and annihilation is performed with a laser beam with spatial Gaussian fluence profile. Like demonstrated in Fig. 3 in the main text, above the lower threshold value $F_{\text{t1}}^{(\dagger)}$ the MSkL ratio $x_{\text{MSkL}}$ scales with the laser fluence until a saturation is reached. Accordingly, the irradiation with a spatial Gaussian fluence profile leads to a formation of a MSkL distribution $\rho_{\text{MSkL}}(r)$ in space. Second, TR-MOKE technique displays the weighted average magnetization dynamics in the probed area. By spatially resolved measurements we average the laser-induced MSkL distribution at different positions. Due to the Gaussian profile of the probe beam, the magnetic structures in the beam center contribute to the TR-MOKE signal to a higher amount than the magnetic structures in the outer region. Consequently, we can describe the evolution of the MSkL ratio with distance quantitatively by the convolution ($\circledast$) of the laser-induced MSkL distribution $\rho_{\text{MSkL}}(r)$ with the normalized Gaussian intensity profile of the probe beam $I(r+d)/\ I_0$ according to 
\begin{equation}
x_{\text{MSkL}}(d) = \rho_{\text{MSkL}}(r) \circledast \frac{I(r+d)}{I_0} \label{eq:convolution}
\end{equation}
with the normalized intensity profile of the probe beam
\begin{equation}
\frac{I(r+d)}{I_0}  = \exp\left( \frac{-2(r+d)^2}{w_L^2} \right)\label{eq:probebeam}
\end{equation}
with beam radius $w_L$. The beam radius describes the distance from the center where the light intensity drops to $1/e^2$. The MSkL distribution is best modeled by a high-order Gaussian function with radius $w_{\text{MSkL}}$ given in Eq. 2 in the main text.
We implement the two dimensional convolution in Eq. \ref{eq:convolution} by a space discretization. For that purpose we define a two-dimensional space grid and two circles on this grid. The first circle (circle 1) represents the MSkL distribution, is centered at $(0,0)$ and has a radius of $r_{\text{MSkL}}= 3w_{\text{MSkL}}$. The second circle (circle 2) models the Gaussian probe beam, has a radius of  $r_{L}= 3w_L$ and is positioned at $(d,0)$. The circle radii are defined in the given way to cover 99.7\% of the MSkL distribution and probe intensity, respectively. We assign the grid points within the circles high-order Gaussian (Eq. 2 in main text) and Gaussian (Eq. \ref{eq:probebeam}) distributed values between 1 and 0 for circle 1 and circle 2, respectively. The two-dimensional convolution in Eq. \ref{eq:convolution} is approximated by calculating the overlap of the two circles on the grid as a function of $d$. The distance-dependent MSkL ratios in Fig. 2 of the main text are extracted by fitting $\rho_0$, $w_{\text{MSkL}}$ and $n$ to the experimental data. In Fig. \ref{fig:FigureS6} we show the fit parameters leading to the best agreement between model and data and corresponding to the solid lines in Fig. 2 in the main text.

\section{Estimation of laser heating}
\begin{figure}
\includegraphics{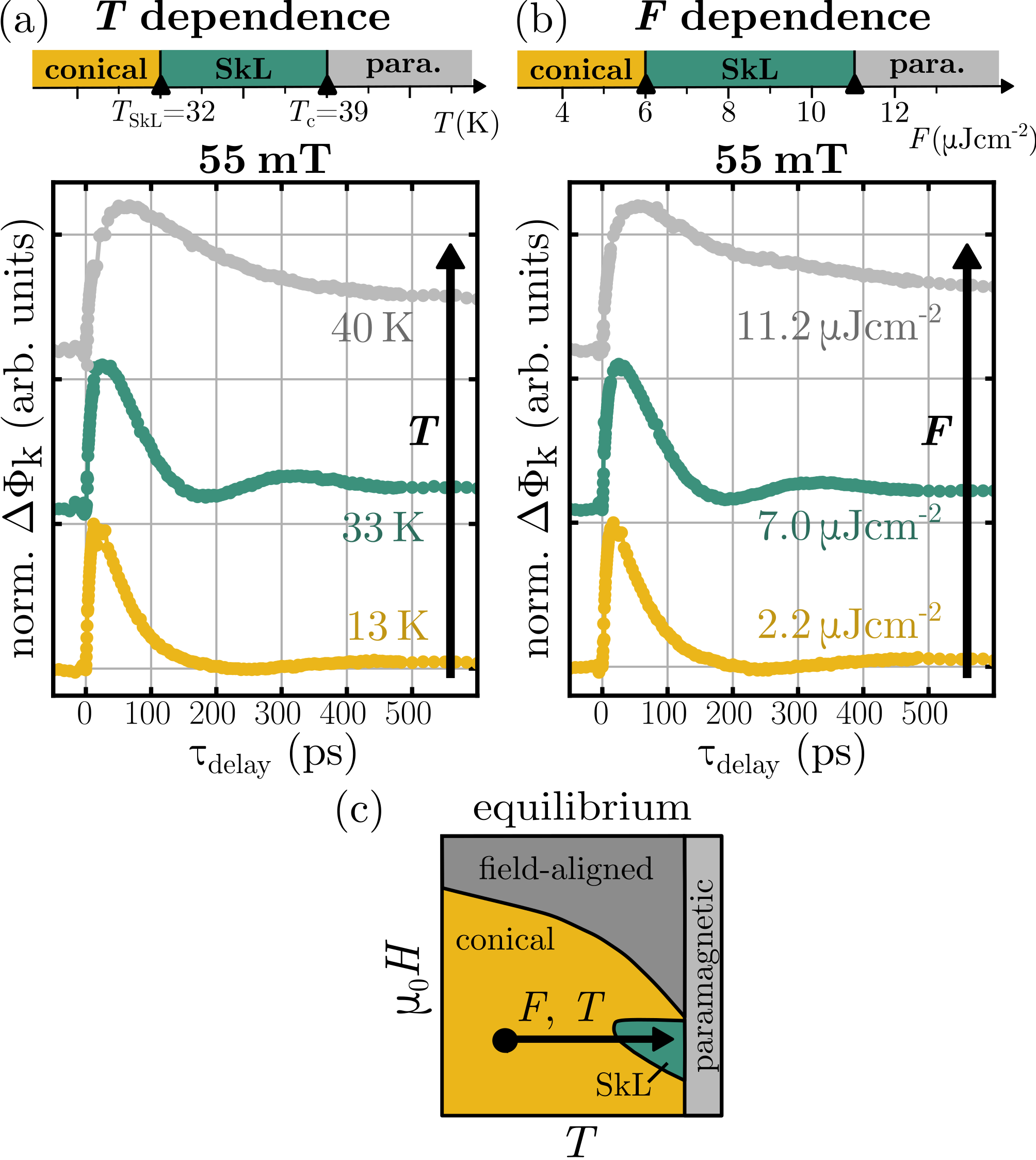}
\caption{\label{fig:FigureS7} (a) Temperature and (b) Pump laser fluence dependent magnetization dynamics of Fe$_{0.75}$Co$_{0.25}$Si at $55\,$mT and $13\,$K in thermal equilibrium. With increasing pump laser fluence we observe the characteristic magnetization dynamics of the conical (yellow), equilibrium SkL (turquoise) and paramagnetic (gray) state. The same dynamics we observe, when increasing the temperature.
}
\end{figure}

\begin{figure}
\includegraphics{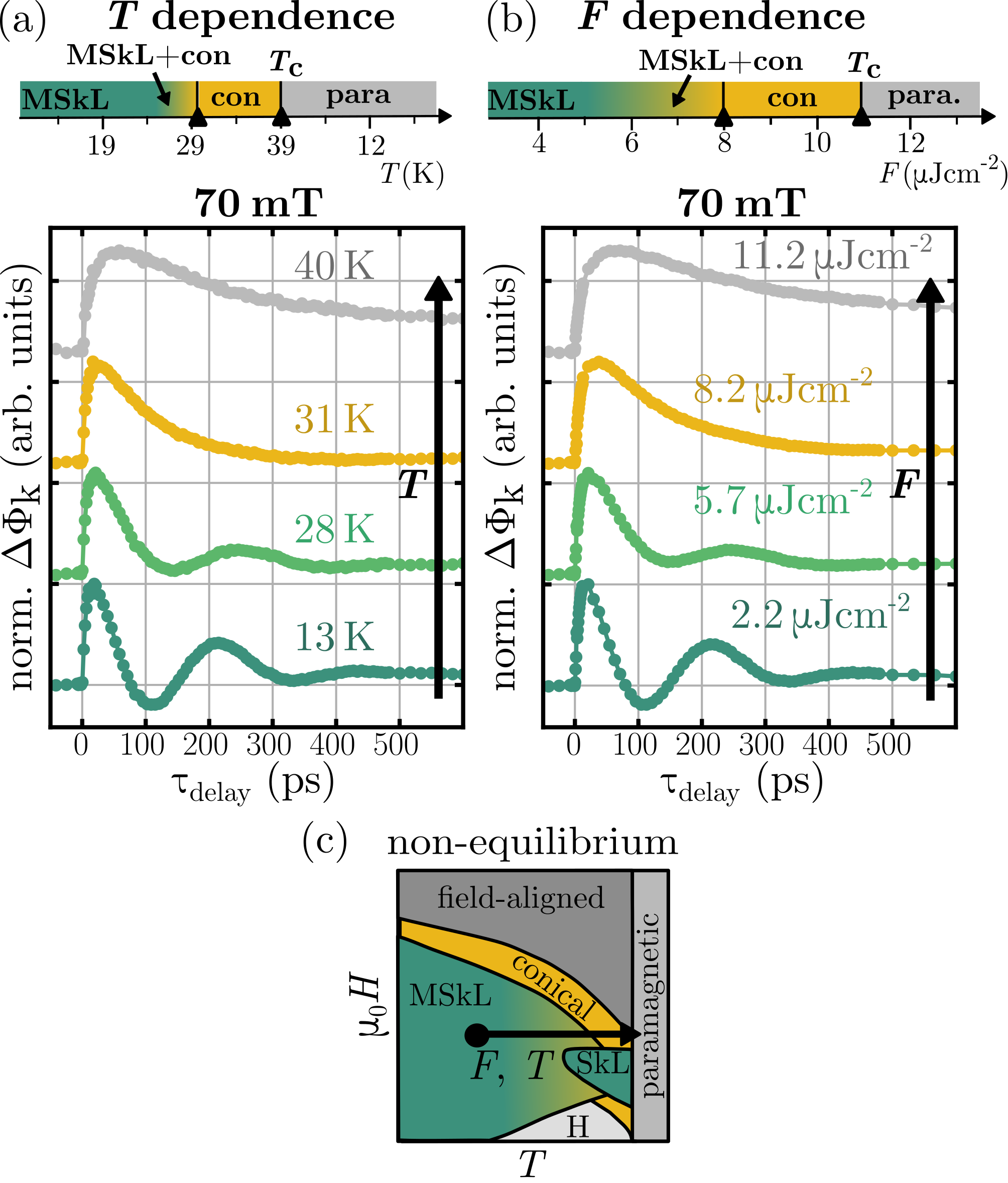}
\caption{\label{fig:FigureS8} (a) Temperature and (b) Pump laser fluence dependent magnetization dynamics of Fe$_{0.75}$Co$_{0.25}$Si in thermal non-equilibrium at $70\,$mT and $13\,$K. With increasing pump laser fluence we observe in a) the characteristic magnetization dynamics of the MSkL (turquoise), the MSkL-conical phase coexistence (green), equilibrium conical (yellow) and paramagnetic (gray) state. The same dynamics we observe, when increasing the temperature.
}
\end{figure}

The irradiation of the sample with focused fs laser pulses leads to a local increase of the sample temperature on different timescales. Next to a transient temperature rise on a sub-ps timescale due to a single fs pulse, the cumulative heat load of multiple laser pulses leads to a local increase of the sample temperature due to steady state heating. 
In order to estimate the temperature increase due to steady state laser heating, we compare temperature- and fluence-dependent magnetization dynamics measured in TR-MOKE experiments. 
For this we prepare the sample in the conical state at $55\,$mT and $13\,$K and either successively increase the sample temperature by cryogenic control or increase the pump laser fluence. In Fig. \ref{fig:FigureS7}(a), TR-MOKE measurements under non-invasive conditions with a pump fluence of $F= 2.2\,\mu$Jcm$^{-2}$ are shown for different sample temperatures. We observe the characteristic magnetization dynamics of the conical, equilibrium and paramagnetic state within well-defined temperature regimes. In comparison, Fig. \ref{fig:FigureS7}(b) shows TR-MOKE measurements for different pump fluences at fixed cryogenic temperature and magnetic field. In dependency of the pump fluence, again the three well-known magnetization dynamics can be observed. Thereby laser fluences smaller than $6\,\mu$Jcm$^{-2}$ prepare the conical state, up to $11\,\mu$Jcm$^{-2}$ the equilibrium SkL state can be identified and for higher fluences, the paramagnetic state is reached. Fig. \ref{fig:FigureS8} repeats the comparison starting in the MSkL state ($70\,$mT, $13\,$K). By cryogenic temperature and pump fluence control, the MSkL state evolves towards the conical state which is finally reached at $T= 31\,$K and $F = 8\,\mu$Jcm$^{-2}$ and transfers into the paramagnetic state at $T= 39\,$K and $F = 11\,\mu$Jcm$^{-2}$. Due to the similar behavior of the magnetization dynamic under cryogenic temperature variation and change of the pump fluence, we explain the pump fluence dependence of the magnetization dynamics by a laser-induced increase of the local temperature due to steady state heating. This allows us to express the irradiation fluence thresholds $F_{t1}^{(\dagger)}$ and $F_{t2}^{(\dagger)}$ in terms of local temperature increase due to laser heating, as shown in Fig. 3(e) and (f) in the main text. 

\section{Finite element modeling of steady state laser heating in Fe$_{0.75}$Co$_{0.25}$Si}
After identifying steady state laser heating as the reason for laser-induced MSkL creation and annihilation in Fe$_{0.75}$Co$_{0.25}$Si, we study the spatial temperature evolution in finite element heat flux modeling with COMSOL [36]. As we are only interested in the cumulative heat load of multiple laser pulses, we simulate the temperature evolution caused by the irradiation with a continuous wave laser with the average laser fluence of our experiment. By that we neglect the ultrafast temperature change caused by a single pulse. Yet, to simplify comparison to the main text, in the following we still refer to the corresponding laser pulse fluence $F$. Temperature dependent bulk material parameters, e.g., thermal conductivity $\kappa$ and heat capacity $C$, were taken from references [27,37] and are listed in Table \ref{tab:coefficients}. The heat transfer coefficient was calibrated by fluence dependent measurements to reproduce the results in Fig. 3(e) and (f) in the main text.

\begin{figure}
\includegraphics{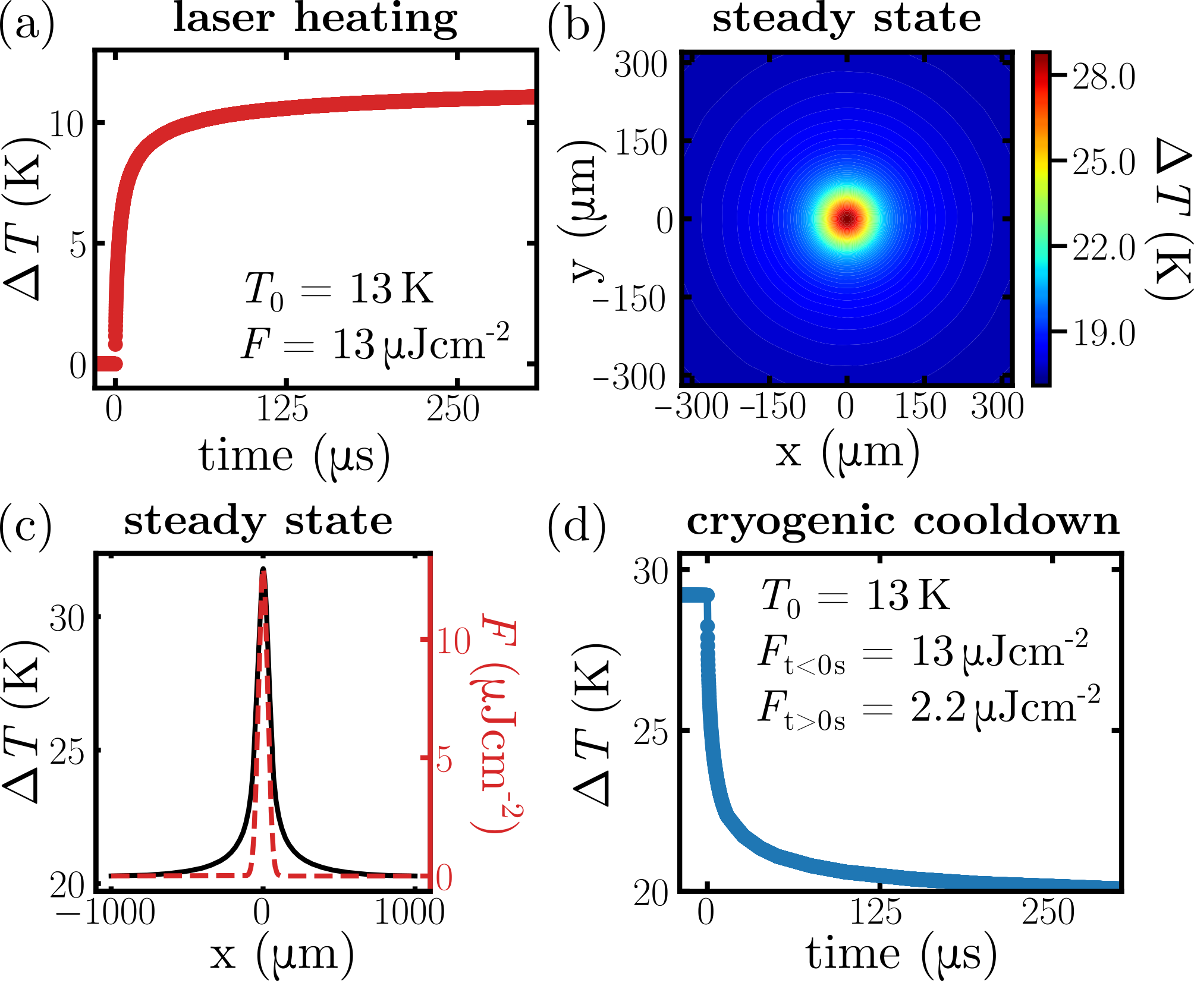}
\caption{\label{fig:FigureS9} Steady state laser heating: (a) Change of the sample temperature $\Delta T$ by laser heating with irradiation time. (b) Temperature distribution at the sample surface in steady state. (c) Comparison of the steady state temperature profile for $y=0\,\mu$m (solid black curve) and the laser fluence profile (dashed red curve). (d) Sample temperature evolution after reducing the laser fluence from $F_{t<0\,\text{s}}=13\,\mu$Jcm$^{-2}$ to $F_{t>0\,\text{s}}=2.2\,\mu$Jcm$^{-2}$.
}
\end{figure}

First, we simulate the sample temperature change in the center of the laser beam due to the irradiation with laser pulses of $F= 13\mu$Jcm$^{-2}$. In Fig. \ref{fig:FigureS9}(a) the sample temperature evolution in the first $300\,\mu$s of laser irradiation are shown for a start temperature of $13\,$K. We observe a sudden increase of the temperature on a sub-$\mu$s-timescale with irradiation time and a subsequent slow variation until a steady state is reached after approximately $1\,$s. This demonstrates that the laser-induced temperature rise is non-linear in time, which might be due to the different thermal reservoirs involved.
The heat map in Fig. \ref{fig:FigureS9}(b) shows the sample surface temperature spatially resolved after a steady state is reached. The laser heating leads to a local increase of the sample temperature, which agrees well with the creation and annihilation of micrometer-sized metastable skyrmion patches. Due to thermal diffusion the laser-induced temperature distribution is much broader than the Gaussian laser beam as shown in Fig. \ref{fig:FigureS9}(c). This explains why the radius of the MSkL distribution shown in the main text in Fig. 2 and in Fig. \ref{fig:FigureS6} is larger than the laser focal radius for some fluences.
Finally, we simulate the cryogenic cooldown of the sample after the laser fluence is reduced. This cooldown leads to the thermal quenching of the SkL, when performed in the field regime of the skyrmion pocket.
In Fig.  \ref{fig:FigureS9}(d) the temperature evolution of the sample in the center of the laser beam is shown after reducing the laser fluence from $F_{t<0\,\text{s}}=13\,\mu$Jcm$^{-2}$ to $F_{t>0\,\text{s}}=2.2\,\mu$Jcm$^{-2}$. The sample is cooled by the thermal bath provided by the cryostat. Again, we observe a fast change of the sample temperature with time and a subsequent slower variation until a steady state is reached (after approximately $1-2\,$s). Also the cooldown subsequent to laser irradiation is non-linear in time and can not be described by a single time constant. 
Overall finite element modeling of steady state laser heating in Fe$_{0.75}$Co$_{0.25}$Si shows the local character of this heating technique, and validates the main findings of laser-induced MSkL creation and annihilation.

\begin{table}[h!]
  \begin{center}
    
    \begin{tabular}{l|c} 
     \textbf{Material parameter} & \textbf{Value }\\
     \hline
	electronic heat capacity $\gamma_e$ & $19\,$mJ mol$^{-1}K^{-2}$ [27]\\
	thermal conductivity $\kappa$ & $20.9\,$Wm$^{-1}$K$^{-1}$[27, 37] \\
	thermal capacity C& $2.5\,$Jmol$^{-1}$K$^{-1}$ [27] \\
	thermal diffusion $D$& 1 cm$^2$/s [37]\\
	penetration depth $\xi$& $30\,$nm\\
    \end{tabular}
\caption{Material parameters of Fe$_{0.75}$Co$_{0.25}$Si used in the finite element modeling}\label{tab:coefficients}
  \end{center}
\end{table}

